\documentclass[aps,preprint]{revtex4}
\usepackage{amsfonts}
\usepackage{amssymb}
\usepackage{subcaption}
\usepackage{amsmath}
\usepackage{graphicx}
\usepackage[font={footnotesize,it}]{caption}
\usepackage{bigints}
\usepackage{latexsym}
\usepackage{enumitem}

\begin{document}

\title{Probing the Lorentz Invariance Violation via Gravitational Lensing and Analytical Eigenmodes of Perturbed Slowly Rotating Bumblebee Black Holes}

\author{Mert Mangut}
\email{mert.mangut@emu.edu.tr}
\affiliation{Department of Physics, Eastern Mediterranean
University, 99628, Famagusta, North Cyprus via Mersin 10, Turkey}
\author{Huriye Gürsel}
\email{huriye.gursel@emu.edu.tr}
\affiliation{Department of Physics, Eastern Mediterranean
University, 99628, Famagusta, North Cyprus via Mersin 10, Turkey}
\author{Sara Kanzi}
\email{sara.kanzi@final.edu.tr}
\affiliation{Faculty of Engineering, Final International University, North Cyprus via Mersin 10, Kyrenia 99370, Turkey}
\author{\.{I}zzet Sakall{\i}}
\email{izzet.sakalli@emu.edu.tr}
\affiliation{AS245, Department of Physics, Eastern Mediterranean
University, 99628, Famagusta, North Cyprus via Mersin 10, Turkey}

\begin{abstract}
The ability of bumblebee gravity models to explain dark energy, which is  the phenomenon responsible for the universe's observed accelerated expansion, is one of their most significant applications. An effect that causes faster expansion can be linked to how much the Lorentz symmetry of our universe is violated. Moreover, since we do not know what generates dark energy, the bumblebee gravity theory seems highly plausible. By utilizing the physical changes happening around a rotating bumblebee black hole (RBBH), we aim to obtain more specific details about the bumblebee black hole's spacetime and our universe. However, as researched in the literature, slow-spinning RBBH (SRBBH) spacetime, which has a higher accuracy, will be considered instead of general RBBH. To this end, we first employ the Rindler--Ishak method (RIM), which enables us to study how light is bent in the vicinity of a gravitational lens. We evaluate the deflection angle of null geodesics in the equatorial plane of the SRBBH spacetime. Then, we use astrophysical data to see the effect of the Lorentz symmetry breaking (LSB) parameter on the bending angle of light for numerous astrophysical stars and black holes. We also acquire the analytical greybody factors (GFs) and quasinormal modes (QNMs) of the SRBBH. Finally, we visualize and discuss the results obtained in the conclusion section.
\end{abstract}

                              
\maketitle

\section{\label{sec:level1}Introduction}

The study of the Lorentz invariance~\cite{Wiesendanger:2018dzw}, a fundamental principle in Einstein's theory of general relativity, has been a topic of great interest in the scientific community. The Lorentz invariance states that the laws of physics are the same for all observers, regardless of the relative motion. In recent years, various experimental and theoretical attempts have been made to test the validity of this principle, and possible violations have been suggested~\cite{Bailey:2006fd,Greenberg:2002uu,Betschart:2008yi,Khodadi:2021owg,Kanzi:2019gtu}. {While we acknowledge that there are many studies~\cite{Khlebnikov:2007ii,Feng:2018gqr, Feng:2015jlj,Khodadi:2022pqh} that discuss the effect of quantum gravity on black holes, our study focuses specifically on the effect of the Lorentz invariance violation (LIV) on the properties of black holes.

The bumblebee gravity theory~\cite{Neves:2022qyb,Delhom:2019wcm} is a modification of general relativity that includes a non-zero scalar field, also known as the ``bumblebee field''. This field allows for the LIV~\cite{Halprin:1999be,Lehnert:2009qv,Torri:2020dec}, and has been proposed as a way to explore possible deviations from the standard model of physics~\cite{ParticleDataGroup:2006fqo,Chan:2006nf}. One of the earliest studies on the bumblebee gravity {theory was introduced as a toy model by Kostelecky et al.~\cite{Kostelecky:2010ze,Bluhm:2004ep,Bailey:2006fd,Bluhm:2007bd} and by Bertolami and Paramos~\cite{Bertolami:2005bh}}, who derived the vacuum solutions of a bumblebee gravity model with vector-induced spontaneous LSB. This was followed by a series of 
 \mbox{papers~\cite{Casana:2017jkc,Li:2020dln,Li:2020wvn,Ding:2019mal,Maluf:2020kgf,Chen:2020qyp,Kanzi:2021cbg,Khodadi:2021owg,Liu:2019mls,Gullu:2020qzu,Gomes:2018oyd,Kanzi:2022vhp,Maluf:2022knd,Jha:2021eww,Ding:2021iwv,Jiang:2021whw,Ovgun:2018xys,Ovgun:2018ran}} that investigated various aspects of bumblebee gravity, including its effects on the early universe and gravitational lensing, its relation to dark energy, its implications for gravitational waves and thermal radiation, and new black hole and wormhole solutions. In particular, in recent years, there has been growing interest in the studies of bumblebee black holes~\cite{Xu:2023xqh}, which are solutions to the field equations of bumblebee gravity. Casana et~al.~\cite{Casana:2017jkc} initially reported an almost Schwarzschild spacetime solution using the bumblebee model, with the bumblebee field having only a radial component. Very recently, \mbox{Xu et~al.~\cite{Xu:2022frb}} significantly extended Casana et~al.'s static solution and discovered spherical solutions with a temporal component of the bumblebee field. The kinetic term seen in the action~\cite{Xu:2022frb} has suggested that the radial component is non-dynamic and can be removed from the field equations. They found two families of solutions, one from a set of two second-order differential equations and another from a set of three. The temporal component of the bumblebee field is crucial for both families of solutions as it dramatically changes the behavior of the metric near the black hole event horizon. Remarkably, it enables the existence of solutions with a non-vanishing $g_{t t}$ at the event horizon. The static bumblebee black hole solutions have been followed by RBBH \mbox{studies~\cite{Ding:2019mal,Ding:2023niy,Ding:2020kfr,Oliveira:2021abg,Kanzi:2021cbg}}, which account for rotation. In those spinning bumblebee gravity models, the gravitational field is described by a metric that includes cross-diagonal terms that are associated with the rotation of the black hole. The field equations are also modified to account for rotation, resulting in a more complex set of equations~\cite{Chen:2020qyp}. However, studies on RBBH have gained momentum in recent years. The first RRBH solution was found in 2019 by \mbox{Ding et~al.~\cite{Ding:2019mal}}, who showed that the bumblebee field can impact the properties of rotating black holes, including their shapes, sizes, and gravitational attraction. This was followed by several other studies that investigated the effects of rotation on bumblebee black holes, including their QNMs, GFs, shadow, superradiant instability, and many other features~\cite{Liu:2019mls,Li:2020dln,Jha:2020pvk,Kanzi:2021cbg,Khodadi:2021owg,Jha:2021eww,Jiang:2021whw,Wang:2021gtd,Khodadi:2022dff,Kanzi:2022vhp,Kuang:2022xjp,Jha:2022bpv,Liu:2022dcn,Khodadi:2023yiw}. These studies have shed new light on the properties and behaviors of bumblebee black holes and their potential implications for our understanding of the universe. On the other hand, gravitational lensing (the bending of light due to the gravitational attraction of massive objects), GFs (factors that describe how the emission spectrum of the black hole differs from a completely uniform black-body spectrum), and QNMs (characteristic ringing patterns produced by black holes after disturbances) are promising methods for probing the LIV effects. 

Gravitational lensing is a fascinating phenomenon in which the path of light is bent by the gravitational force of a massive object, such as a galaxy or a cluster of galaxies. This effect was first predicted by Einstein's theory of general relativity in 1915, and it has since been observed and studied by astronomers around the world~\cite{Turner:1990mk,Bartelmann:1999yn,Sauer}. Gravitational lensing allows us to study objects that are otherwise too distant or faint to observe directly, as the lensing effect can magnify and distort their appearance. Moreover, by analyzing the way in which the light is bent and distorted, we can learn about the properties of the lensing object and the distribution of matter in the universe. Gravitational lensing has, therefore, become an important tool for studying astrophysics and cosmology, providing us with valuable insights into the nature of the universe and the objects within it.
For computing gravitational lensing, there are various methods~\cite{Bartelmann:2010fz}. RIM~\cite{Rindler:2007zz,Sultana:2012zz} is a particular method used for calculating the bending of light by a black hole. This method is based on the Rindler approximation, which describes the spacetime around a black hole in a simplified way. The RIM also considers the effect of the black hole's rotation on the trajectory of light and calculates the amount of bending that occurs as a result. This method is widely used to study the properties of rotating black holes and has been applied to a variety of problems in astrophysics and cosmology~\cite{Ishak:2008zc,Bhattacharya:2009rv,Bhattacharya:2010xh,Sultana:2012zz,Cattani:2013dla,Heydari-Fard:2014pwa,Fernando:2014rsa,Ali:2017ofu,Secuk:2019svg,Heydari-Fard:2020sib,Mangut:2021suk,He:2020eah}. 

In this article, we aim to explore the potential of probing LIV through the gravitational lensing, GFs, and QNMs of RBBHs. Meanwhile, in black hole physics, a GF~\cite{Maldacena:1996ix,Cvetic:1997xv,Harmark:2007jy,Sakalli:2022xrb,Sakalli:2016abx,Sakalli:2016fif} (also known as an absorption coefficient or a transmission coefficient) is the measure of how much of an incoming wave is absorbed by a black hole and how much is scattered or transmitted away. When a wave (such as a photon or a particle) approaches a black hole, it can be partially absorbed by the black hole's gravitational field, and the remaining energy can be scattered or transmitted away. The probability of absorption depends on the properties of the wave (such as its energy, frequency, and angular momentum) and the properties of the black hole (such as its mass, charge, and spin). A black hole's GF is the ratio of the absorbed flux to the incident flux of the wave. It is called a ``greybody'' factor because it represents a partial absorption of the wave, as opposed to a complete absorption (which would result in a blackbody spectrum). Namely, GFs are important in black hole physics because they affect the thermal radiation emitted by a black hole (known as Hawking's radiation~\cite{Hawking:1975vcx,Hawking:1974rv,Bardeen:1973gs}). The Hawking radiation spectrum is related to the GFs of the black hole and can provide information about the properties of the black hole. In summary, GFs are important in the study of black hole thermodynamics and in the search for black hole candidates in astrophysics. In this context, it is one of the important aims of this study to explore the fingerprints of the LIV parameter with GFs with the semi-analytical bound method~\cite{Sakalli:2022xrb,Ngampitipan:2012dq,Boonserm:2019mon,Boonserm:2009zba,Boonserm:2008dk}, which is a powerful technique used in black hole physics to compute the GFs. The semi-analytical bound method (the so-called Miller--Good transformation method~\cite{Boonserm:2008dk}) involves combining analytical and numerical techniques to calculate the GFs. The analytical part of the method involves finding a set of bounds on the GFs, based on the properties of the black hole and the radiation being emitted. These bounds can be computed using the perturbation theory. One advantage of the semi-analytical bound method is that it can be applied to a wide range of black hole geometries and radiation types, including scalar, electromagnetic, and gravitational radiation. It is also computationally efficient, making it well-suited for studying the properties of black holes in astrophysical scenarios, such as accretion disks and binary systems~\cite{Oshita:2022pkc}. {In short, we shall employ the RIM for the gravitational lensing phenomenon, the Miller--Good transformation method for the GF computations, and the unstable circular null geodesic method, whose results are in agreement with the WKB approximation method~\cite{Cardoso:2008bp,Ovgun:2021ttv}, for the derivation of analytical QNMs. As a result, we aim to shed new light on the nature of LIV and its implications for our understanding of the universe. In the meantime, it can be questioned as to why these different subjects are discussed in the same article. We should clarify this issue as follows: Many theories and models have been proposed to explain gravity, each with its own unique features and characteristics. Among the various theories, the bumblebee gravity theory has been studied extensively in recent years. In the bumblebee gravity theory, the vector field responsible for carrying the gravitational force has an additional symmetry that is not present in the GR theory. This symmetry, called gauge symmetry, is a mathematical property that describes the symmetry of the field under certain transformations. One of the important aspects of gravity is its ability to cause gravitational lensing. Gravitational lensing occurs when light passing through a gravitational field is bent, creating a distortion of the image of the object being observed. This effect has been observed in many astronomical phenomena, such as galaxy clusters and black holes. Hence, the bumblebee gravity theory is an interesting area of research in the field of astrophysics. For this reason, it can be studied in conjunction with other important concepts, such as gravitational lensing, GFs, and QNMs, to gain a deeper understanding of the behavior of bumblebee gravity and its interactions with matter. The study of these concepts and their interrelationships have the potential to yield many exciting discoveries about the bumblebee gravity theory in the years to come.}

This paper is organized as follows. Section \ref{sec1} presents a brief review of SRBBH spacetime and its physical features. In Section \ref{sec2}, we study the gravitational lensing of SRBBH and discuss the results obtained by comparing the real astrophysical data of stars. In Section \ref{sec3n}, {we use the Miller--Good transformation method to compute the GFs of the SRBBH. Then, in Section \ref{sec3}, for a dilatory RBBH spacetime, we derive the analytical QNMs via the unstable circular null geodesic method. Finally, we draw our conclusions in \mbox{Section \ref{sec4}}.}

En passant, unless otherwise stated, we use geometrized (natural) units $G=c=\hbar=k_{B}=1$ and $(-,+,+,+)$ metric signature in this paper.

\section{RBBH Spacetime} \label{sec1}

In this section, we will provide a brief overview of the Einstein--bumblebee gravity model and its associated black hole solutions. This model is an example of an extension to the standard general relativity framework. By utilizing the appropriate potential, the bumblebee vector field $B_\mu$ obtains a non-zero vacuum expectation value, resulting in the breaking of Lorentz symmetry within the gravitational sector. The action for a single bumblebee field $B_\mu$ that is coupled to gravity can be defined as follows~\cite{Casana:2017jkc,Chen:2020qyp}
\begin{align}\label{Action}
\mathcal{S}=&\int d^4x \sqrt{-g}\left[ \frac{1}{2\kappa}\left( \mathcal{R}+\varrho B^aB^b \mathcal{R}_{ab} \right) -\frac{1}{4}\left(B^{ab}B_{ab}+4\mathcal{V}\right) \right]+\mathcal{L}_M,
\end{align}
where $ \kappa=8\pi G_N $, {$\mathcal{R}_{ab}$ is the Ricci tensor, and $\mathcal{R}$ denotes the Ricci scalar.} The actual strength of the connection between non-minimal gravity and the bumblebee field is determined by the real coupling constant,  which is denoted as $ \varrho $. The magnitude of the bumblebee field $B_{a b}$ is specified by a definition of its intensity:
\begin{align}
B_{a b}=\partial_a B_b-\partial_b B_a.
\end{align}

The bumblebee field $B_a$ has a vacuum expectation value that should not be zero, so a potential $\mathcal{V}$ is selected accordingly. The potential {$\mathcal{V}$} has a minimum at $y_{0}=B_aB^a\pm b^2$ where $b$ is a real, positive constant. Therefore, the potential can be represented using the formula given in the 2018 paper by Casana~\cite{Casana:2017jkc}.
\begin{align}
\mathcal{V}=\mathcal{V}(y_{0}).
\end{align}

The possibility of a non-zero vacuum value $\langle B^a\rangle=b^a$ exists and it can be triggered by a certain potential, which satisfies the condition: $b_ab^a=\mp b^2$. The process of calculating the variation in action \eqref{Action} leads to the formulation of two equations: one describes the behavior of gravity in an empty space (known as the vacuum gravitational equation), and the other describes how the bumblebee field moves (known as the equation of motion for the bumblebee field). Hence, one has
\begin{align} \label{tr1c}
& \mathcal{R}_{ab}-\frac{1}{2}g_{ab}\mathcal{R}=\kappa T^{B}_{ab}, \\
& \nabla^aB_{ab}=2{\mathcal{V}}'B_b-\frac{\varrho}{\kappa}B^a\mathcal{R}_{ab}. \label{tr11c}
\end{align}
where $T^{B}_{ab}$ denotes the bumblebee energy-momentum tensor~\cite{Ding:2019mal}:

\begin{equation}
\begin{aligned}
& T_{ab}^B=B_{ac} B_a^c-\frac{1}{4} g_{ab} B^{cd} B_{cd}-g_{ab} \mathcal{V}+2 B_a B_b \mathcal{V}^{\prime}+\frac{\varrho}{\kappa}\left[\frac{1}{2} g_{ab} B^c B^d \mathcal{R}_{cd}-B_a B^c \mathcal{R}_{cb}-B_b B^c \mathcal{R}_{ca}\right. \\
& \left.+\frac{1}{2} \nabla_c \nabla_a\left(B^c B_b\right)+\frac{1}{2} \nabla_c \nabla_b\left(B^c B_a\right)-\frac{1}{2} \nabla^2\left(B_a B_b\right)-\frac{1}{2} g_{ab} \nabla_c \nabla_d\left(B^c B^d\right)\right], \label{izr0}
\end{aligned}
\end{equation}

\vspace{-9pt}
Moreover, ${\mathcal{V}}'=\partial {\mathcal{V}}(y)/\partial y$ at $y=y_{0}$. With the trace of Equation \eqref{tr11c}, one obtains the trace-reversed version:
\begin{align} \label{izr01}
\mathcal{R}_{ab}=&\kappa T^{B}_{ab}+2\kappa g_{ab}{\mathcal{V}}-\kappa g_{ab}B^cB_c{\mathcal{V}}'+\frac{\varrho}{4}g_{ab}\left(\nabla^2\left(B^cB_c \right)+2g_{ab}\nabla_c\nabla_d(B^cB^d)\right).
\end{align}

{Now, without loss of generality, one can assume that the bumblebee field is frozen at its vacuum expectation value. This assumption makes the specific form of the potential, which is irrelevant to its dynamics, resulting in $V=0$ and $V^{\prime}=0$. With these conditions, the first two terms in Equation \eqref{izr0} are similar to those of the electromagnetic field, except for the coupling terms to the Ricci tensor. Given this, Equation \eqref{izr01} leads to the equations for the gravitational field~\cite{Ding:2019mal}: $R_{ab}=0$, with
\begin{align}
R_{ab}=&\mathcal{R}_{ab}-\kappa b_{ac}b^{c}_{b}+\frac{\kappa}{4}g_{ab}b^{cd}b_{cd}+\varrho b_ab^c\mathcal{R}_{cb}+\varrho b_bb^c\mathcal{R}_{ca}-\frac{\varrho}{2}g_{ab}b^cb^d\mathcal{R}_{cd}+\bar{\mathcal{B}}_{ab}, \label{izr}
\end{align}
\vspace{-15pt}
\begin{align}
\bar{\mathcal{B}}_{ab}=&-\frac{\varrho}{2}\left[\nabla_c \nabla_a\left(b^c b_b\right)+\nabla_c \nabla_b\left(b^c b_a\right)-\nabla^2\left(b_a b_b\right)\right]. \label{izr1}
\end{align}

Overall, the condition of $R_{ab}=0$ determines whether the obtained spacetime is an exact solution of the vacuum Einstein--bumblebee action or not. Additional elaborations regarding this matter were presented in a very recent study by Liu et al~\cite{Liu:2022dcn}}.

{In 2020, Ding et~al.~\cite{Ding:2019mal} revealed that by imposing the condition $b^ab_a=const.$ and utilizing the bumblebee field $b_a=(0,\rho b_0\Delta^{1/2},0,0)$, a Kerr-like black hole solution for the Einstein--bumblebee theory can be obtained. This solution is expressed as follows:}

\begin{equation}
ds^2=-\left(1-\frac{2M r}{\rho^2}\right)dt^2+\rho^2\left(\frac{1}{\Delta}dr^2+d\theta^2\right)+\frac{1}{\rho^2}\left(-4M r \tilde{a}\sin^2\theta dtd\varphi+\mathcal{A}\sin^2\theta d\varphi^2\right). \label{izr2}
\end{equation}

where
\vspace{-6pt}
\begin{align}
\mathcal{A}=&\left(r^2+\tilde{a}^2 \right)^2-\Delta\tilde{a}^2\sin^2\theta,\\
\rho^2=&r^2+\tilde{a}^2\cos^2\theta, \\
\Delta=&(1+\ell)^{-1}(r^2-2Mr)+a^2, \\
\tilde{a}=&\sqrt{1+\ell}a.
\end{align}

\vspace{-6pt}
The Lorentz-violating parameter (we shall also call it the {bumblebee or LSB parameter~\cite{Tuleganova:2023izp}}) in metric \eqref{izr2} is denoted by $\ell=\varrho b_0^2$, and $a$ is the rotation parameter; this metric represents a rotating spacetime with a radial bumblebee field. When $\ell\rightarrow 0$, the metric reduces to the usual Kerr metric, while it becomes the static Einstein--bumblebee metric when $a\rightarrow 0$. {However, the metric given in Equation \eqref{izr2} is not a correct RBBH solution according to the findings of Maluf and Muniz~\cite{Maluf:2022knd}; their work was supported by~\cite{Liu:2022dcn,Kanzi:2021cbg,Kanzi:2022vhp}. In particular, it was shown in~\cite{Maluf:2022knd} that
\begin{align}
 \nabla^\theta b_{\theta r}+\frac{\varrho}{\kappa} b^r \mathcal{R}_{r r}&\neq0,\\
  \nabla^r b_{r\theta}+\frac{\varrho}{\kappa} b^r \mathcal{R}_{r \theta}&\neq0.
\end{align}

However, it was again emphasized by Maluf and Muniz~\cite{Maluf:2022knd} that metric \eqref{izr2} becomes correct in the slow rotation limit.} To address this issue, we can expand the metric to the second order of the rotation parameter $\tilde{a}$ by assuming that $\tilde{a}$ is sufficiently small, and we obtain the following metric:
\begin{multline}\label{izr3}
ds^2\approx-\left(1-\frac{2M}{r}+\frac{2M\tilde{a}^2\cos^2\theta}{ r^3}\right)dt^2 
-\frac{4M\tilde{a}\sin^2\theta}{r}dtd\varphi+\frac{(1+\ell)\rho^2}{\tilde{\Delta}}dr^2 \\ 
+\left(r^2+\tilde{a}^2\cos^2\theta\right)d\theta^2+\sin^2\theta\left[r^2+\tilde{a}^2\left(1+\frac{2M}{r}\sin^2\theta\right) \right]d \varphi^2,
\end{multline}

In the second order of $\tilde{a}$, $\tilde{\Delta}$ is the same as $\Delta$. This enables us to identify the horizons $r_{+}$ (event) and $r_{-}$ (inner) by finding the roots of $\tilde{\Delta}$.
\begin{align}
\tilde{\Delta}=(r-r_{+})(r-r_{-}).
\end{align}
where
\begin{align}\label{rh}
r_{+}=2M-r_{p},~~~~r_{-}=r_{p},
\end{align}
in which
\begin{align}
r_{p}=\frac{\tilde{a}^2}{2M}.  
\end{align}

Metric \eqref{izr3} still does not fully satisfy certain components of the field Equation \eqref{izr}. On the other hand, the non-zero terms for the field equations are all linked to the rotation parameter $a$. Therefore, metric \eqref{izr3} can be viewed as the slow 
} RBBH (SRBBH) estimation of the Einstein--bumblebee equation: $\bar{R}_{ab}=0$. More information about this issue can be found in the Appendix section of Reference~\cite{Liu:2022dcn}. For the slowest or dilatory RBBH (DRBBH), one can consider the case of $\tilde{a}^2 \rightarrow 0$ and hence $r_{+}\rightarrow2M$ and $r_{-}=r_{p}\rightarrow0$. In this case, metric \eqref{izr3} becomes
\begin{equation}\label{izr3n}
ds^2\approx-f(r)dt^2-\frac{2r_{+}\tilde{a}\sin^2\theta}{r}dtd\varphi+g(r)dr^2+r^2\left(d\theta^2+\sin^2\theta d\varphi^2\right),
\end{equation}
where 
\begin{align}
 f(r)=&1-\frac{r_{+}}{r},\\
 g(r)=&\frac{(1+\ell)}{f(r)}.
\end{align}

At this stage, to avoid lengthy computations and prioritize clarity and comprehensibility, let us discuss the thermodynamic features of the DRBBH. To this end, we first consider a particle's behavior in close proximity to the event horizon by utilizing the following four-velocity~\cite{Schutz:1985jx}:
\begin{equation}
u^\alpha \rightarrow\left(\left(-g^{t t}\right)^{1 / 2}, 0, 0,\frac{-g^{t \varphi}}{\left(-g^{t t}\right)^{1 / 2}}\right).
\end{equation}

The zero angular momentum observer (ZAMO) always behaves well because $g^{t t}<0$ for all cases outside the horizon $\left(r>r_{+}\right)$. The ZAMO always co-rotates with the black hole at the following angular velocity (as seen from spatial infinity):

\begin{equation}
\Omega_{\text {ZAMO }}=\frac{u^\varphi}{u^t}=-\frac{g^{t \varphi}}{g^{t t}}=-\frac{g_{t \varphi}}{g_{\phi \varphi}}  
\end{equation}

One may ask at a given $(r, \theta)$ what the range of the allowed $\mathrm{d} \varphi / \mathrm{d} t$ is. One can judge that this is possible by the following metric condition:
\begin{equation}
g_{t t}+2 g_{t \varphi} \frac{\mathrm{d} \varphi}{\mathrm{d} t}+g_{\varphi \varphi}\left(\frac{\mathrm{d} \phi}{\mathrm{d} t}\right)^2<0,
\end{equation}
or $\Omega_{\min }<\frac{\mathrm{d} \varphi}{\mathrm{d} t}<\Omega_{\max }$ with
\begin{equation}
\Omega_{\max , \min }=\Omega_{\text {ZAMO }} \pm \frac{\sqrt{\left(g_{t \varphi}\right)^2-g_{t t} g_{\varphi \varphi}}}{g_{\varphi \varphi}}.
\end{equation}

What is interesting is that as we approach the event horizon $r_{+}$, although $g_{\phi \phi}$ remains finite, $\left(\left(g_{t \varphi}\right)^2-g_{t t} g_{\varphi \varphi}\right) \rightarrow 0:$; thus, all particles near the event horizon must have
\begin{equation} \label{zamo}
\frac{\mathrm{d} \phi}{\mathrm{d} t} \approx \Omega_{\mathrm{H}}=\Omega_{\mathrm{ZAMO}}\left(r_{+}\right)=\frac{a}{r_{+}^2}.
\end{equation}

Thus, $\Omega_{\mathrm{H}}$ is called the angular velocity of the horizon or the black hole. It is obvious that the RBBH metric is stationary and axisymmetric, with Killing fields $\xi^a=(\partial / \partial t)^a$ and $\psi^a=(\partial / \partial \phi)^a$. Moreover, 
 the RBBH is asymptotically flat, which can be seen crudely from the fact that the metric components \eqref{izr3} or \eqref{izr3n} approach those of the Minkowski spacetime in spherical polar coordinates as $r \rightarrow \infty$. Employing the black hole mass definition of Wald~\cite{Wald:1984rg}
\begin{equation}\label{izm1}
\tilde{M}=-\frac{1}{8 \pi} \int_S \epsilon_{a b c d} \nabla^c \xi^d,
\end{equation}
where the integral is taken over a sphere, $S$, one can obtain the mass of the DRBBH
\begin{equation} \label{izm2}
\tilde{M}=\frac{r_{+}}{2 \sqrt{1+l}}=\frac{M}{\sqrt{1+l}}.
\end{equation}

Moreover, the total angular momentum of RBBH can be computed via the following expression~\cite{Wald:1984rg}: 
\begin{equation} \label{izm3}
J=\frac{1}{16 \pi} \int_S \epsilon_{a b c d} \nabla^c \psi^d,
\end{equation}
which yields the following expression for the DRBBH
\begin{equation} \label{izm4}
J=\frac{\tilde{a}r_{+}}{2 \sqrt{1+l}}=\frac{ar_{+}}{2}=Ma.
\end{equation}

As the metric components solely rely on $r$ and $\theta$, the acceleration of a particle can be derived from
\begin{equation} 
\mathit{a}^{\alpha}=\Gamma _{\mu \nu }^{\alpha}u^{\mu }u^{\nu
}=-g^{\alpha \mu }\partial _{\mu }\ln u^{t}.  \label{izr14}
\end{equation}

According to Reference~\cite{Wald:1984rg}, the surface gravity ($\kappa $) is defined as
\begin{equation} 
\kappa =\lim_{r\rightarrow r_{+}}\sqrt{\mathit{a}^{\alpha }%
\mathit{a}_{\alpha}}(u^{t})^{-1}.  \label{izr15}
\end{equation}

As an exemplary study, if we apply the above formulation of the surface gravity \big(i.e., Equation \eqref{izr15}\big) to metric \eqref{izr3n}, and make some straightforward calculations, we obtain the surface gravity of the DRBBH as follows
\begin{equation} 
\kappa =\frac{1}{2}\left.  \partial _{r} f\right\vert _{r=r_{+}}=\frac{1}{2r_{+}\sqrt{1+\ell}}.  \label{izr16}
\end{equation}

Thus, the Hawking temperature of the DRBBH reads
\begin{equation} 
T_{H}=\frac{\kappa }{2\pi }=\frac{1}{4\pi r_{+}\sqrt{1+\ell}}.  \label{izr17}
\end{equation}

The black hole area is given by
\begin{equation} 
A_{BH}=\int_{0}^{2\pi }d\varphi \int_{0}^{\pi }\sqrt{-g} d\theta =4\pi r_{+}^2.  \label{izr19}
\end{equation}%

Therefore, one can easily derive the entropy of the DRBBH as follows
\begin{equation} 
S_{BH}=\frac{A_{BH}}{4\hslash }=\pi r_{+}^2.  \label{izr21}
\end{equation}

The obtained thermodynamical quantities of the DRBBH, which are given in Equations \eqref{zamo}, \eqref{izm2}, \eqref{izm4}, \eqref{izr17}, and \eqref{izr21} imply that the first law of thermodynamics~\cite{Bardeen:1973gs} 
\begin{align} \label{izm11}
d\tilde{M}&=T_{H}dS_{BH}+\Omega _{H}\ dJ, \\ \nonumber 
&=\frac{dr_{+}}{2\sqrt{1+l}}+\frac{a^2dr_{+}}{2r_{+}^2},
\label{22}
\end{align}
holds when $(\frac{a}{M})^{2}\rightarrow 0$. Thus, both the nearly static DRBBHs and/or the significantly massive DRBBHs satisfy the first law of thermodynamics.  

\section{Gravitational Lensing of SRBBH via RIM} \label{sec2}
\begin{figure}[!ht]
\includegraphics[width=10cm]{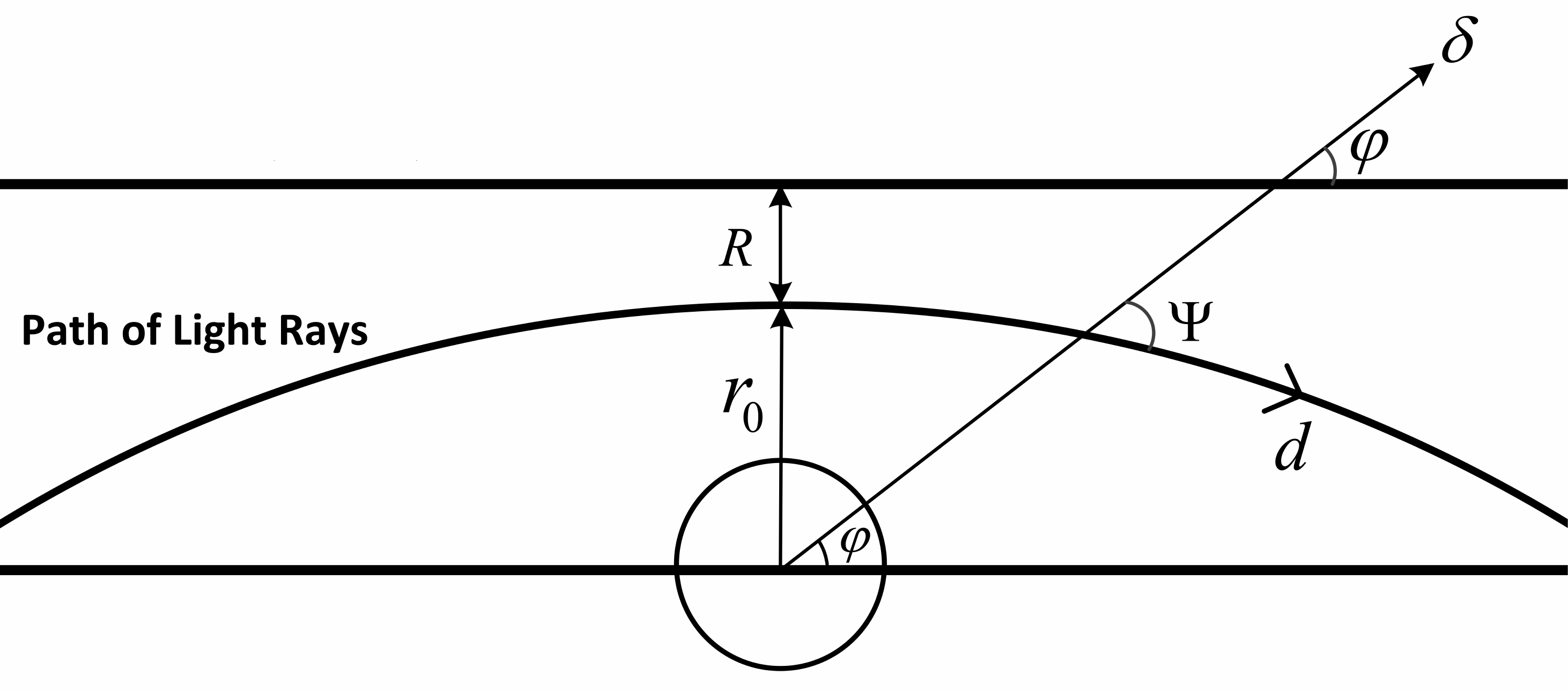} 
\caption{The one-sided bending angle is given by $\epsilon =\psi-\varphi .$ The upper straight line symbolizes the undistorted light rays defined by the solution of the homogeneous part of Equation \eqref{mr1}.} \label{figx1}
\end{figure}

Gravitational lensing is a phenomenon in black hole physics where the intense gravitational field of a black hole bends and distorts the path of light passing near it. This can result in the creation of multiple images of the same source, or even the formation of a complete ring of light known as the Einstein ring. In the vicinity of a black hole, the strong gravitational field warps spacetime, causing the paths of light rays to curve. This can lead to an effect where a distant object appears distorted or magnified when viewed from a certain angle. This phenomenon is similar to how a magnifying glass bends and focuses light to make objects appear larger. Gravitational lensing has been observed in a variety of astrophysical contexts, including around supermassive black holes at the centers of galaxies, and around smaller black holes in binary systems. In some cases, it has even been used to indirectly detect the presence of black holes themselves, by observing the effects of their gravity on the light from nearby stars or gas. In this section, we shall analyze the gravitational lensing of SRBBH via RIM~\cite{Rindler:2007zz}. This method is based on the invariance of the angle, which is obtained as a result of the scalar product of two vectors:

\begin{equation}
\cos \left( \psi \right) =\frac{d^{i}\delta _{i}}{\sqrt{\left(
d^{i}d_{i}\right) \left( \delta ^{j}\delta _{j}\right) }}
\end{equation}
where { the vector directions $d$ and $\delta$ } can be represented geometrically in the standard symmetry plane {($\theta=\pi/2$ plane)}, with $t=const.$, as shown in Figure~\ref{figx1}.
\newpage
The geometrical polar coordinates of the directions can be written as 
\begin{eqnarray}
d &=&\left( dr,d\varphi \right) =\left( A,1\right) d\varphi \text{ \ \ \ \ \
\ \ }d\varphi <0,  \notag \\
\delta &=&\left( \delta r,0\right) =\left( 1,0\right) \delta r,
\end{eqnarray}
where $A(r,\varphi)=\frac{dr}{d\varphi}$. By mapping metric \eqref{izr3n} to the following general rotating spacetime: 
\begin{equation}
ds^{2}=-f(r)dt^{2}-2g(r)dtd\varphi+h(r)dr^{2}+p(r)d\varphi^{2},
\end{equation}
in which
\begin{align} \label{izss1}
    f(r)=&1-\frac{2M}{r}, \notag\\
    g(r)=&\frac{2M\tilde{a}}{r}, \\
    h(r)=&\frac{(\ell+1)^2}{1-\frac{2M}{r}+\frac{\tilde{a}}{r^2}}, \notag\\
    p(r)=&r^2+\tilde{a}^2\left(1-\frac{2M}{r}\right) \notag,
\end{align}

One can obtain the following null geodesic equation~\cite{mert} { through the standard change of variable process, $u=\frac{1}{r}$ (recall the classical Kepler problems),} and subsequently derive the evolution of the light curve in the SRBBH spacetime \eqref{izr3n}: 
\begin{equation}
\frac{d^{2}u}{d\varphi^{2}}=\frac{1}{2}\frac{d}{du}\left[\frac{u^{4}\left(f(u)p(u)+g^{2}(u)\right)\left(p(u)-2g(u)b-f(u)b^{2}\right)}{h^{2}(u)\left(g(u)+f(u)b\right)^{2}}\right], \label{Mert2}
\end{equation}
where $E$ and $L$ are the energy and angular momentum of the photon, respectively, and the parameter defined as $b=E/L$ is called the impact parameter. If we put the metric function into Equation~\eqref{Mert2}, the generic null geodesic equation becomes
\begin{multline}
\frac{d^{2}u}{d\varphi ^{2}}+\beta u\simeq\frac{M u^2 \left(-3 a^2 b (\ell+1)-2 a \sqrt{\ell+1} \left(9 a^2 (\ell+1)+12 M^2\right)+3 b^3\right)}{b^3 (\ell+1)^2}-\\
\frac{2 a M}{b^3 (\ell+1)^{3/2}}, \label{mr1}
\end{multline}
and 
\begin{equation}
\beta=\frac{\left(-3 a^2 b (\ell+1)+8 a \sqrt{\ell+1} M^2+b^3\right)}b^3 (\ell+1)^2.
\end{equation}

If we use the standard approximation of $u=\frac{\text{sin}\left[\sqrt{\beta} \varphi\right]}{b}$ to solve Equation \eqref{mr1} and put the homogeneous  solution into the inhomogeneous side of Equation \eqref{mr1}, then the first order perturbative solution can be found as follows
\begin{equation}
 u(\varphi)\approx\frac{\text{sin}\left[\sqrt{\beta} \varphi\right]}{b}+\frac{M \left(\cos \left(2 \sqrt{\beta} \varphi\right)+3\right)}{2 b^2 \beta (\ell+1)^2}-\frac{2 a M}{b^3 \beta (\ell+1)^{3/2}}+O\left(\frac{1}{b}\right)^4. \label{mr4}
\end{equation}

The closed distance approach $r_{0}$ is analyzed at $\varphi =\pi /2$. Thus, one can obtain the closed distance as follows
\begin{equation}
\frac{1}{r_{0}}=\frac{\sin\left[\sqrt{\beta} \pi/2\right]}{b}+\frac{M \left(\cos \left( \sqrt{\beta} \pi\right)+3\right)}{2 b^2 \beta (\ell+1)^2}-\frac{2 a M}{b^3 \beta (\ell+1)^{3/2}}.
\end{equation}

The invariant formula of RIM for the rotating spacetimes is given by~\cite{mert}
\begin{equation}
\tan \left( \Psi \right) =\frac{\left[ h^{-1}(r)p(r)\right] ^{1/2}}{%
\left\vert A(r,\varphi )\right\vert }.  \label{mert3}
\end{equation}

When we substitute Equation \eqref{mr4} into the definition of $A(r,\varphi)$, we can find

\begin{equation}
A(r,\varphi)=-\frac{ \left(\cos \left(\sqrt{\beta} \varphi\right) \left(b \beta (\ell+1)^2-2 M \sin \left(\sqrt{\beta} \varphi\right)\right)\right)}{b^2 \sqrt{\beta} (\ell+1)^2}r^2. \label{mr5}
\end{equation}

If we use the standard approximations of the RIM, $\varphi =0$ and $\frac{M}{R}<<1$, Equations \eqref{mr4} and \eqref{mr5} then yield
\begin{equation}
r\approx \frac{b^2 \beta (\ell+1)^2}{2M},\text{ \ \ \ \ \ } A(r,\varphi=0)%
\approx- r^{2}\frac{\sqrt{\beta}}{b}.     \label{mert6}
\end{equation}

When we use expression \eqref{mert6} and the related metric functions, Equation \eqref{izss1} in Equation \eqref{mert3}, and perform the small angle approximation, $\epsilon=tan\psi_{0}\approx\psi_{0}$, the bending angle of the SRBBH 
can be found as follows:
\begin{equation}
 \begin{aligned}
\epsilon \approx \frac{2M}{\beta^{3/2} (\ell+1)^{3}b}&\left\{1-\frac{2M^{2}}{\beta(1+\ell)^2b^2}+\frac{4M^{2}a^{2}}{\beta^2(1+\ell)^3b^4}-\frac{16M^{4}a^{2}}{\beta^3(1+\ell)^5b^6} \right.\\
 &\left. +\frac{8 M^2\left(a^4 (\ell+1) M^2+4 a^2 M^4\right)}{ \beta^4 (\ell+1)^7b^8}\right\}+O\left(\frac{1}{b}\right)^9.\label{7}
 \end{aligned} 
\end{equation}

\begin{figure}[!ht]
  \begin{tabular}{@{}cccc@{}}
    \includegraphics[width=.28\textwidth]{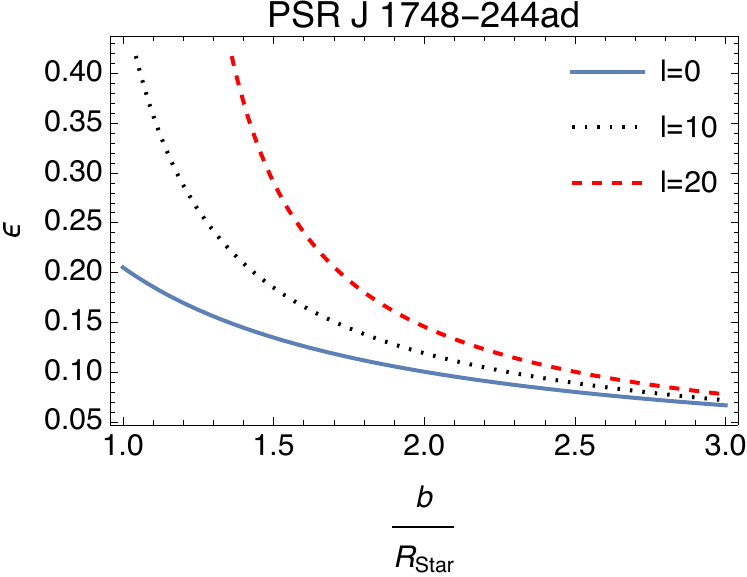} &
    \includegraphics[width=.28\textwidth]{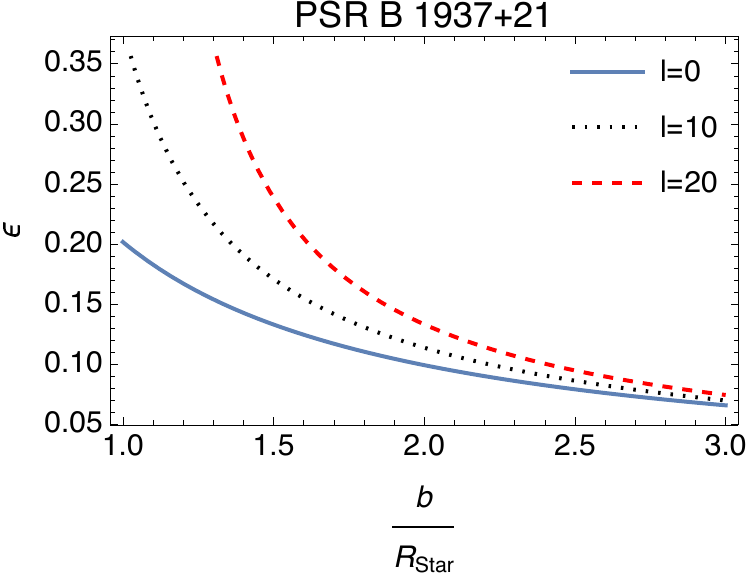} &
  
                                                          \\
   \includegraphics[width=.28\textwidth]{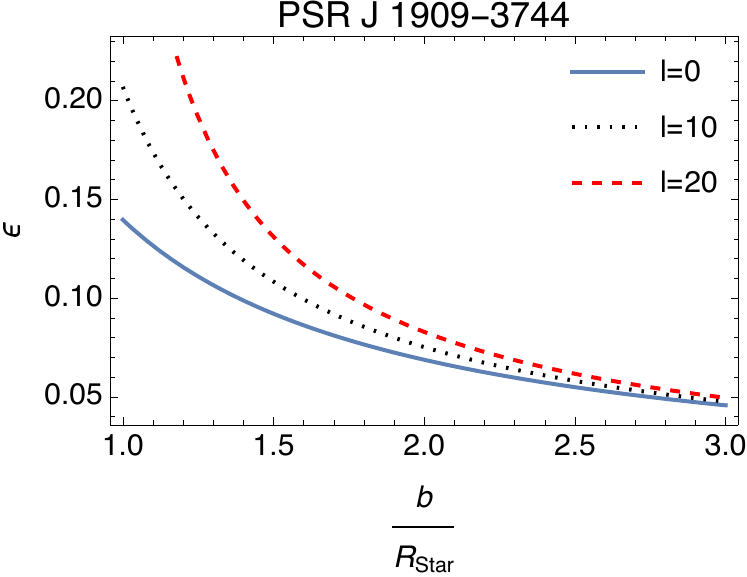} &
    \includegraphics[width=.28\textwidth]{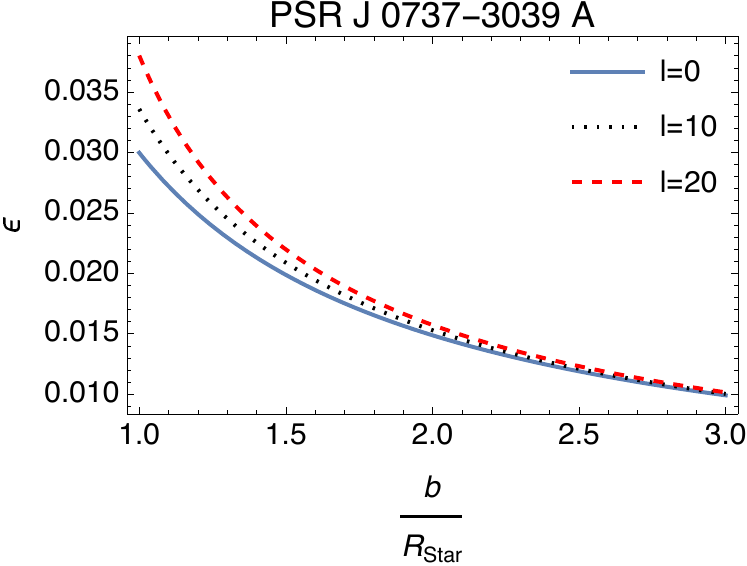} &
                                                             \\
    \multicolumn{2}{c}{\includegraphics[width=.28\textwidth]{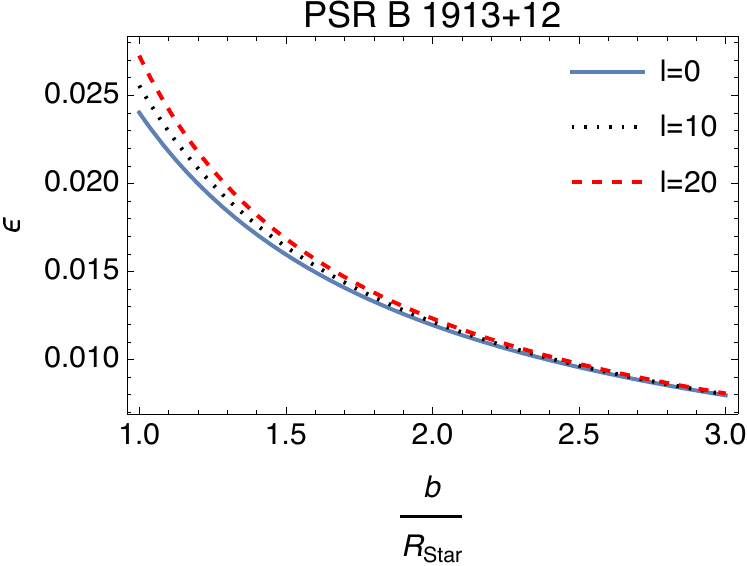}}
  \end{tabular}
  \caption{Bending angle $\epsilon$ versus $\frac{b}R_{star}$ graphs. The $R_{star}$ represents the radius of the star entitled in each graph. The effects of varying bumblebee parameters on the bending angle are illustrated.} \label{fig2}
\end{figure}

We analyzed the gravitational lensing phenomenon for stars whose masses, radii, and rotational parameters are recorded in Reference~\cite{mert}, for specific values of the $\ell$ parameter, by utilizing the graphs presented in Figure~\ref{fig2}. The analysis of the plots depicted in Figure~\ref{fig2} for various real stars observed in the cosmos demonstrates a clear correlation between the bumblebee parameter $\ell$ and the bending angle at low $\frac{b}{R_{star}}$ values. It is worth noting that the units used in the calculation are first converted to standard international units, or S.I units. In order to convert the mass ($M$) to S.I units, it is multiplied by $Gc^{-2}$, where $G=6.67408\times
10^{-11}~$m$^{3}$kg$^{-1}$s$^{-2}$ is the 
 gravitational constant and $c=3\times
10^{8}$~ms$^{-1}$ is the speed of light. This results in the one-sided bending angle being measured in radians. Specifically, as the value of $\ell$ increases, the bending angle also increases. This observation suggests that the bumblebee parameter plays a crucial role in determining the gravitational lensing effects of stars. Furthermore, this trend is only observed at low $\frac{b}{R_{star}}$ values, which indicates that the impact of the bumblebee parameter may be negligible under certain conditions. Overall, these findings provide valuable insights into the underlying physics of gravitational lensing and offer a potential avenue for further investigation into the nature of the bumblebee parameter and its role in gravitational lensing.

\newpage

\section{GFs of SRBBH Spacetime}\label{sec3n}
The calculation of GFs is an important aspect of studying black hole thermodynamics and Hawking's radiation. According to Hawking's famous calculation~\cite{Hawking:1975vcx,Hawking:1974rv}, black holes emit radiation due to quantum effects, and radiation has a thermal spectrum with a temperature proportional to the black hole's surface gravity. However, this calculation assumes that the black hole is a blackbody, namely a perfect absorber, which is not the case in reality. The GF takes into account the deviation from perfect absorption and modifies the thermal spectrum accordingly. In this section, our aim is to derive the GFs of the SRBBH spacetime. In this regard, we use the massive Klein--Gordon equation, which plays an important role in the study of black hole physics. It is a relativistic wave equation that describes the behavior of a scalar field in the presence of a massive particle. In the context of black holes, this equation is often used to describe the behavior of matter fields around black holes. In general, the massive Klein--Gordon equation can be written as~\cite{Kanzi:2019gtu}:
\begin{equation}
    \frac{1}{\sqrt{-g}}\partial_{\mu}(\sqrt{-g}g^{\mu\nu}\partial_{\nu}\phi)=\mu^{2}\phi, \label{s1}
\end{equation}
where $\mu$ indicates the mass of the scalar field. {By substituting metric \eqref{izr3n} in Equation \eqref{s1} with the following ansatz:
\begin{equation}
  \phi=\frac{\Psi(r)}{\sqrt{r^2+a^2}}e^{-i\omega t}Y(\theta,\phi), \label{s2}
\end{equation}
and, in the sequel, expanding the result up to the first order of $\Tilde{a}$, we have as follows (for a similar analysis, the reader is referred to~\cite{Liu:2022dcn} and references therein):} 
\begin{equation}
    \frac{d^2}{dx^2}\Psi_{l}^{(1)}+V_{l}^{(1)}\Psi_{l}^{(1)}=0, \label{s3}
\end{equation}
where $dr/dx=F=1-\frac{2M}{r}$. The effective potential $V_{l}^{(1)}$ is given by 
\begin{equation}
    V_{l}^{(1)}=(1+\ell)\left(\omega^2-\sqrt{1+\ell}\frac{4amM\omega}{r^3}\right)-F\left(\frac{2M}{r^3}+(1+\ell)(\frac{\lambda}{r^2}+\mu^2)\right). \label{s4}
\end{equation}

In Equation \eqref{s4}, $\lambda=l(l+1)$ is the eigenvalue of the wave equation and $l$ stands for the angular quantum number. 
Moreover, the second-order expansion of Equation \eqref{s1} with respect to $\Tilde{a}$ recasts in
\begin{equation}
    \frac{d^2}{dr_{*}^2}Z_{l}+V_{l}^{(2)}Z_{l}=0, \label{s5}
\end{equation}
where $\frac{dr}{dr_{*}}=\frac{(1+\ell)\Tilde{\delta}}{r^2+a^2}$ and
\begin{equation}
    Z_{1}=\Psi_{l}^{(2)}+a^2c_{l}\Psi_{l-2}^{(2)}-a^2c_{l+2}\Psi_{l+2}^{(2)},
\end{equation}
in which 
\begin{equation}
    c_{l}=\frac{(1+l)(\mu^2-\omega^2)}{2(2l-1)}\sqrt{\frac{(l-1)^2-m^2}{4(l-1)^2-1}}\sqrt{\frac{l^2-m^2}{4l^2-1}}.
\end{equation}

Furthermore, the effective potential in Equation \eqref{s5} can be found as follows
\begin{multline}
     V_{l}^{(2)}=V_{l}^{(1)}-\frac{24M^2a^2}{r^6}+\frac{2Ma^2}{r^5}(6-2l(l+1)(1+\ell)-3\ell-2r^2(1+\ell)\mu^2+r^2(1+\ell)^2\omega^2)\\
    +\frac{a^2}{r^4}(l-1+2\ell-l\ell^2+m^2(1+\ell)^2-l^2(\ell^2-1)+r^2\mu^2-r^2(\omega^2+\ell^2\mu^2-\ell^2\omega^2))\\-\frac{a^2F}{r^2}(1+\ell)^2(\mu^2-\omega^2)(\frac{l^2-m^2}{4l^2-1})(\frac{(l+1)^2-m^2}{4(l+1)^2-1}).\label{s6}
\end{multline}

\begin{figure}[!ht]
  \includegraphics[width=.49\textwidth]{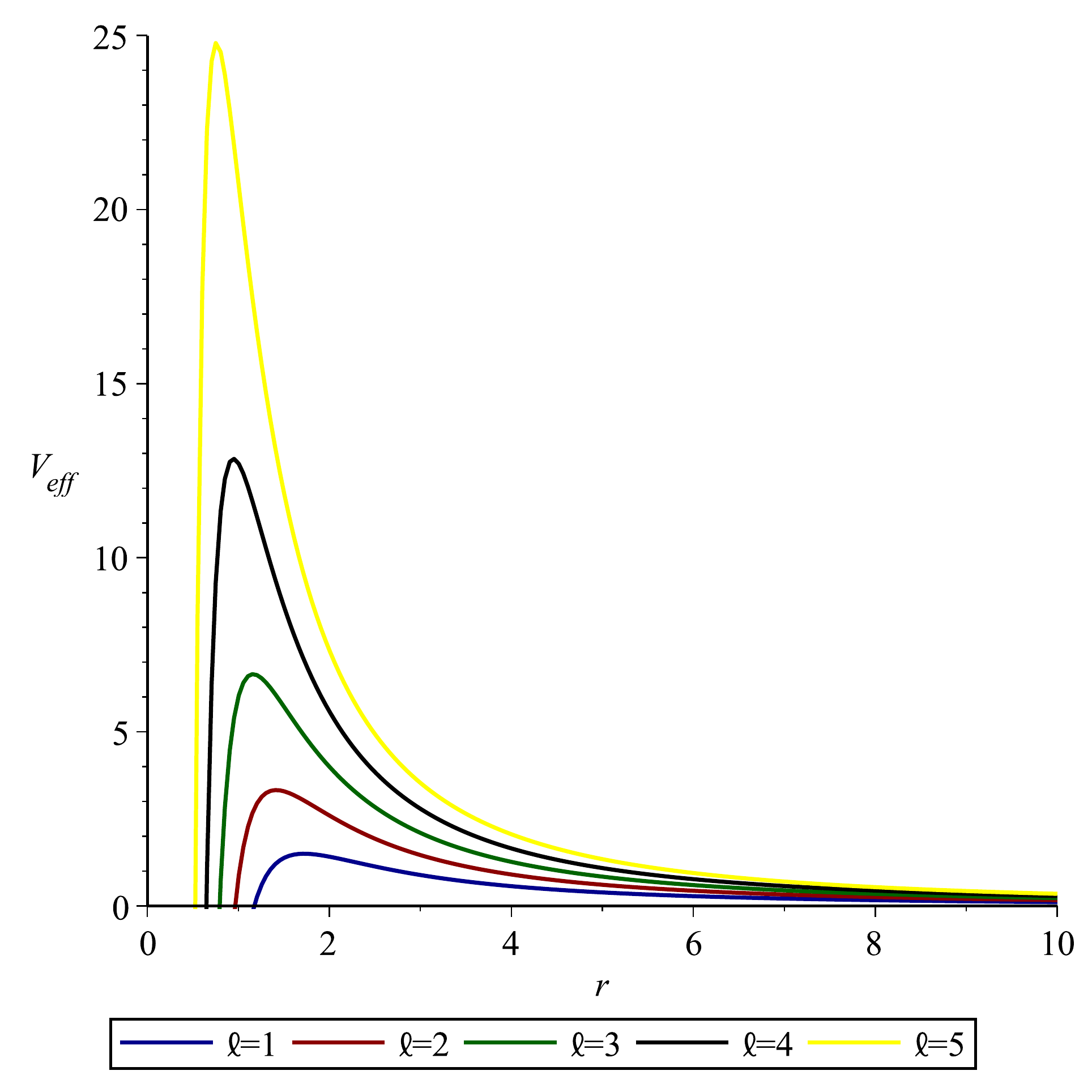}
  \includegraphics[width=.49\textwidth]{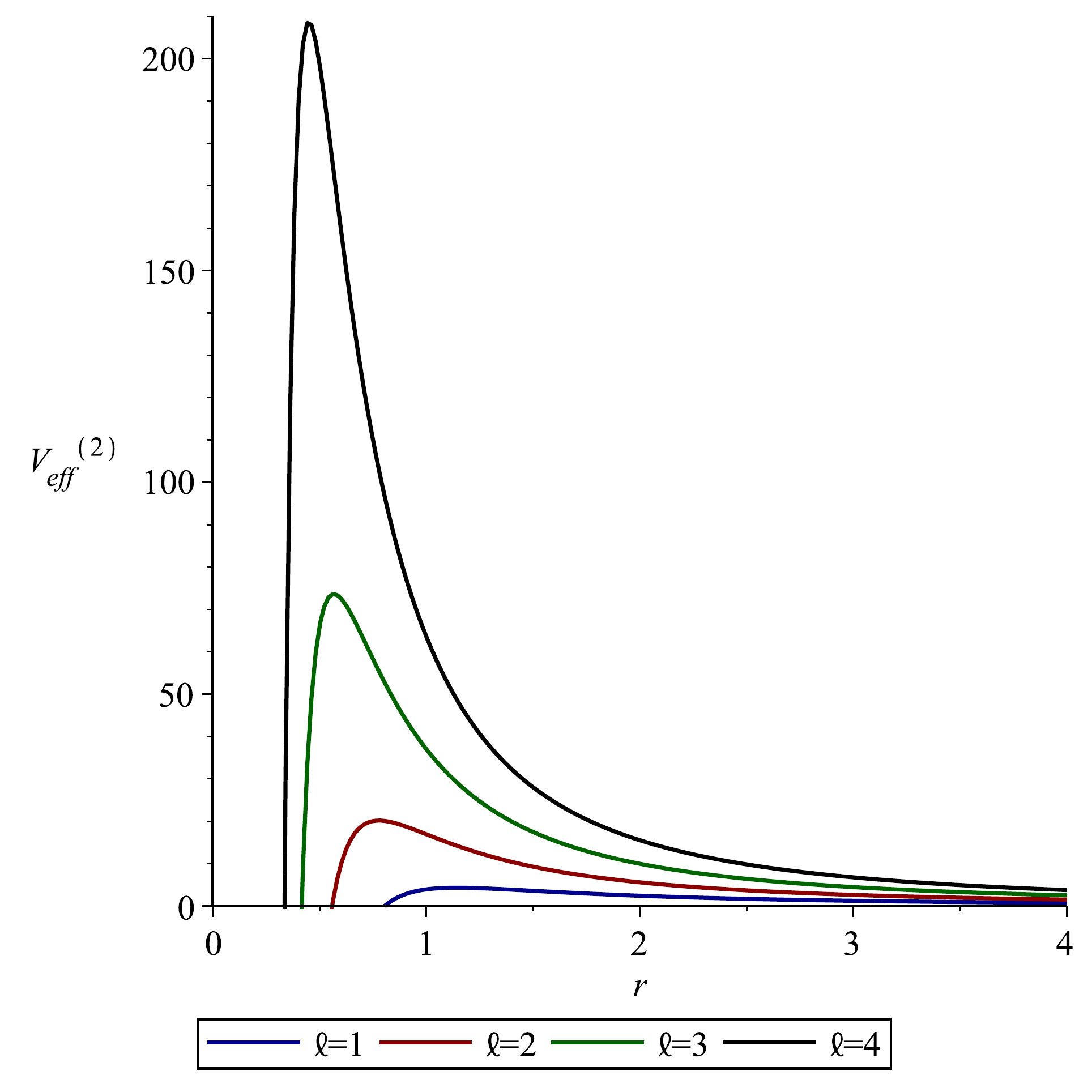}
\caption{$\sigma_{l}(\omega)$ versus $\omega$ graphs of the effective potential for the first order (\textbf{left}) and second order (\textbf{right}) of $\Tilde{a}$. The physical parameters are chosen as $M=m=1, \Tilde{a}=0.1, \lambda=6  \: (l=2)$, and $\omega=10$. } \label{fig:13a}
\end{figure}

  Figure \ref{fig:13a} shows the effective potentials represented in Equations \eqref{s4} and \eqref{s6}. It is evident that the effective potentials increase as a barrier with the increase of the bumblebee parameter $\ell$.

To derive the GFs, we apply the general semi-analytic bounds, defined by

\begin{equation}
\begin{split}
\sigma(\omega) \geq \sec h^2\left[ \int_{-\infty}^{+\infty} \wp dr_*\right]
\end{split}\label{0S1},
\end{equation}
where
\begin{equation}
\begin{split}
\wp=\frac{\sqrt{(h'^2)+(\omega^2-V_{eff}-h^2)^2}}{2h}
\end{split}, \label{Sara1}
\end{equation}

 In which $h(x)$, seen in the integrand of Equation \eqref{0S1}, is a positive function that fulfills the two conditions $: (1) \:h(r_*)>0$ and $(2)\: h(-\infty) = h(+ \infty) = \omega $. After applying the aforementioned conditions to the effective potentials, one may observe a direct proportionality between the GFs and the effective potential, where the metric function plays a significant action in this process. Thus, Equation \eqref{0S1} becomes
\begin{equation}
    \sigma(\omega) \geq \sec h^2\left[ \int_{r_{h}}^{+\infty} \frac{V_{eff}}{2\omega } \frac{dr}{dr_{*}}\right]. \label{26}
\end{equation}

One advantage of using a massless scalar field is that it allows for the propagation of waves at the speed of light, which is a fundamental feature of many physical systems. Additionally, massless fields are mathematically simpler to work with and can lead to more elegant and concise solutions to certain problems. For this reason, by taking cognizance of massless fields and substituting Equation \eqref{s4} in Equation \eqref{26} with $\frac{dr}{dr_{*}} \equiv \frac{dr}{dx}$, one can obtain the GFs of the SRBBH for the first order of $\Tilde{a}$ as follows
\vspace{-6pt}
\begin{multline}
    \sigma_{\ell}^{(1)}(\omega)\geq\sec h^2\bigg(\frac{1}{2\omega}\bigg[4aMm\omega((1+\ell)\sqrt{1+\ell})\big(\frac{1}{2r_{h}^2}+\frac{2M}{3r_{h}^3}+\frac{M^2}{r_{h}^4}+\frac{8M^3}{5r_{h}^5}\big)+\\
    \frac{M+r_{h}\lambda(1+\ell)}{r_{h}^2}\bigg]\bigg).
\end{multline}

\begin{figure}[h]
\includegraphics[width=9cm,height=10cm]{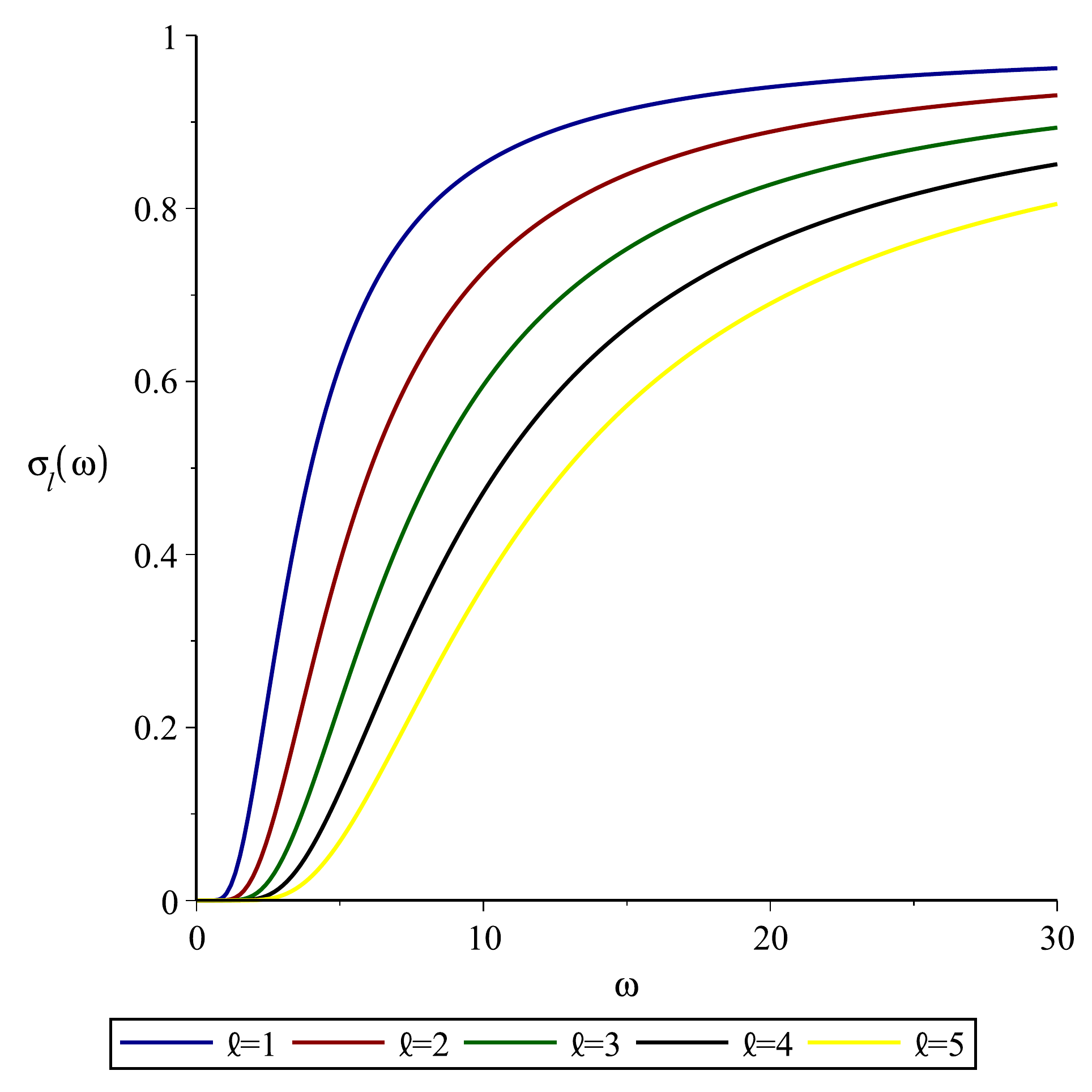}\caption{$\sigma_{l}(\omega)$ versus $\omega$ graphs for the massless scalar field propagating in the SRBBH with the first-order expansion of $\Tilde{a}$. The physical parameters are chosen as $M=1, \lambda=6$, and $\Tilde{a}=0.1$}  \label{SF4}
\end{figure}

The behavior of the GFs for the first order of the rotating parameter $\Tilde{a}$ for the varying bumblebee parameter $\ell$ is depicted in Figure~\ref{SF4}. The figure demonstrates that increasing the bumblebee parameter $\ell$ results in a decreased probability of encountering \mbox{Hawking radiations.}

\begin{figure}[!ht]
  \includegraphics[width=.49\textwidth]{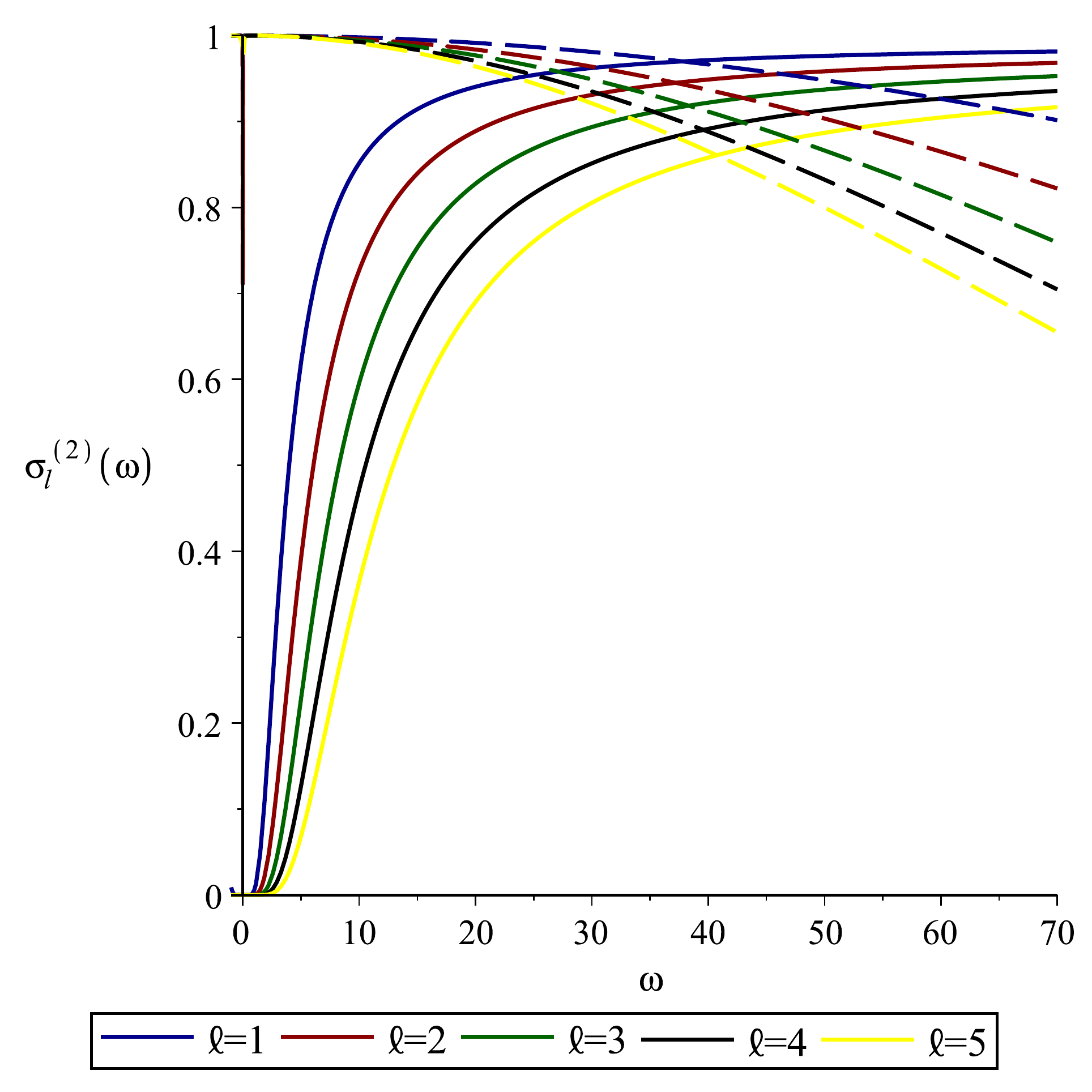}
  \includegraphics[width=.49\textwidth]{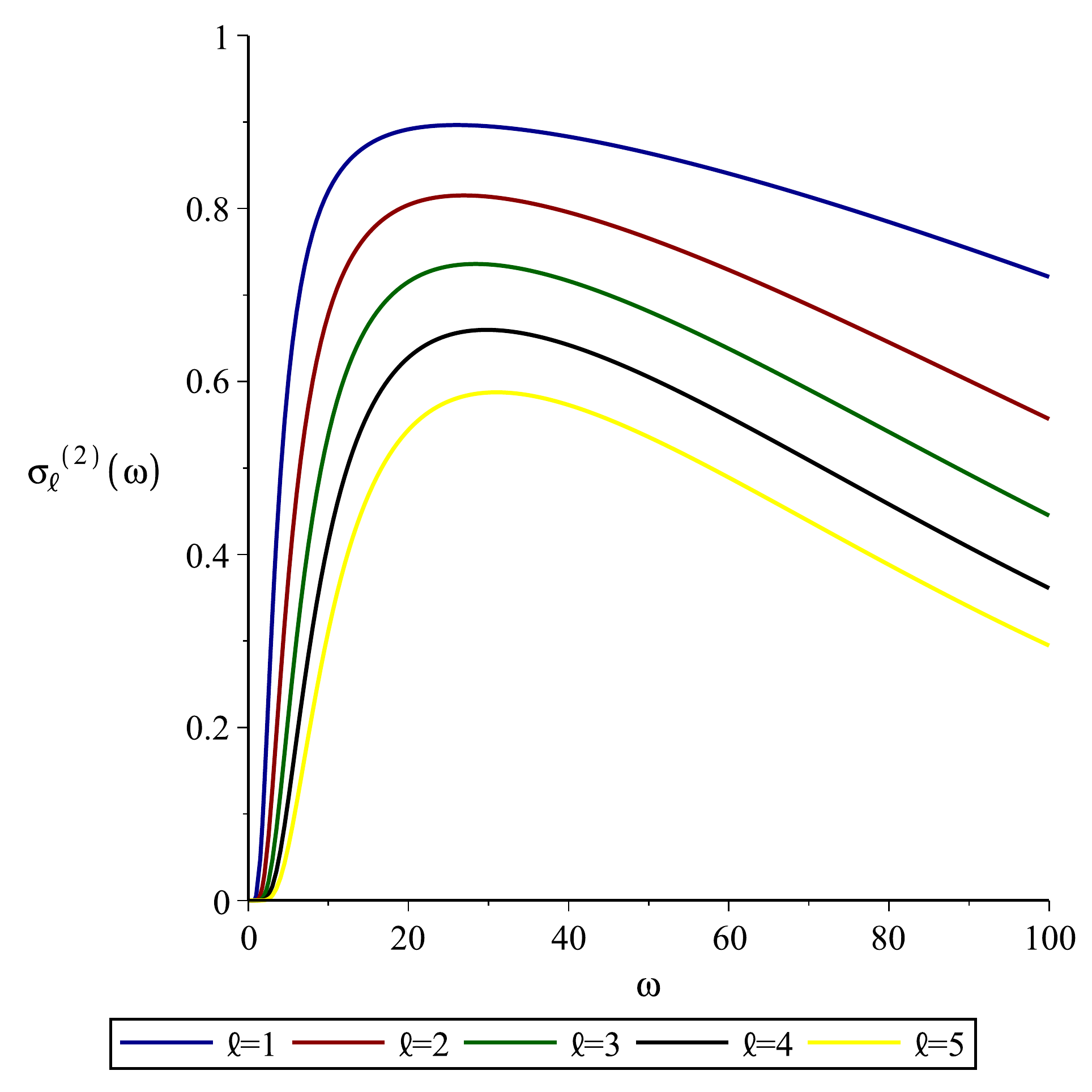}
\caption{$\sigma_{l}(\omega)$ versus the $\omega$ graph for the massless scalar field and first-order expansion of $\Tilde{a}$ for RBBH. The left figure represents the GFs of SRBBH for both the first order (solid lines) and second order (dashed line) of $\Tilde{a}$. The right one shows the GFs of the SRBBH with the combination of both the first and second orders of $\Tilde{a}$; see Equation \eqref{s8}.  } \label{fig:14a}
\end{figure}

The GFs for the second order of $\Tilde{a}$ can be determined using a similar procedure by substituting Equation \eqref{s6} into Equation \eqref{26}. One can obtain

\begin{equation}
    \sigma_{\ell}(\omega)^{(2)}\geq \left[\sigma_{\ell}^{(1)}(\omega)+\sec h^2\left(\frac{1}{2\omega(1+\ell)}\left[\frac{-A}{7r_{h}^7}+\frac{B}{6r_{h}^6}+\frac{C}{5r_{h}^5}+\frac{D}{4r_{h}^4}+\frac{E}{3r_{h}^3}+\frac{H}{2r_{h}^2}\right]\right)\right], \label{s8}
\end{equation}

by which
\begin{equation}
    A=24M^2a^2(a^2-r_{+}r_{-}+4M^2),
\end{equation}
\begin{equation}
    B=2Ma^2(a^2-r_{+}r_{-}+4M^2)(6-2l(l+1)(1+\ell)+3\ell)-48M^3a^2,
\end{equation}
\begin{multline}
    C=a^2(a^2-r_{+}r_{-}+4M^2)(l-1+2\ell-\ell^2l+m^2(1+\ell)^2-l^2(\ell^2-1))\\+4M^2a^2(3\ell-2l(l+1)(1+\ell)),
\end{multline}
\vspace{-29pt}

\begin{multline}
    D=2Ma^2(1+\ell)^2\omega^2(a^2-r_{+}r_{-}+4M^2)+2Ma^2(l-1+2\ell-l\ell^2+m^2(1+\ell)^2-l^2(\ell^2-1))\\+2Ma^2(6-2l(l+1)(1+\ell)+3\ell),
\end{multline}
\vspace{-21pt}
\begin{multline}
    E=Fa^2(1+\ell)^2\omega^2(a^2-r_{+}r_{-}+4M^2)(\frac{l^2-m^2}{4l^2-1}+\frac{(l+1)^2-m^2}{4(l+1)^2-1})-a^2\omega^2(a^2-r_{+}r_{-}+4M^2)\\ \times (1+\ell^2)+4M^2a^2(1+\ell)^2\omega^2+a^2(l-1+2\ell-l\ell^2+m^2(1+\ell)^2-l^2(l^2-1)),
\end{multline}
\begin{equation}
H=2Ma^2(1+\ell)^2\omega^2-2Ma^2\omega^2(1+\ell^2)+2FMa^2(1+\ell)^2\omega^2(\frac{l^2-m^2}{4l^2-1}+\frac{(l+1)^2-m^2}{4(l+1)^2-1}),
\end{equation}

where $r_{+}r_{-}=(2M-(1+\ell)\frac{a^2}{2M})(1+\ell)\frac{a^2}{2M}$. Figure~\ref{fig:14a} represents the relationship between the GFs of the second order of $\Tilde{a}$ and the bumblebee parameters. The right figure is drawn for both the first order $\sigma_{\ell}(\omega)$ and second order $\sigma_{\ell}^{(2)}(\omega)$; however, the left figure represents the GFs of the SRBBH together with the first and second  orders of $\sigma_{\ell}(\omega)$, i.e., Equation \eqref{s8}. Again, it is obvious that the increase in the $\ ell$ parameter decreases the GFs.

\section{QNMs of Extremely Slow Rotating DRBBH via Unstable Circular Null \mbox{Geodesic Method}}\label{sec3}
Ever since the unveiling of the LIGO experiment~\cite{Fricke:2011dv}, there has been an increased interest in gravitational waves, specifically those emitted by disturbed black holes. These waves are predominantly characterized by ``quasinormal ringing''~\cite{Kokkotas:1999bd}, which refers to the damped oscillations at specific frequencies that are unique to the system in question. These QNM frequencies can be determined by calculating the scalar perturbation of a massless field around a black hole:

\begin{equation} 
\partial_{\mu}\left(\sqrt{-g} g^{\mu v} \partial_{\nu} \Psi\right)=0. \label{iz1}
\end{equation}

In the extreme slow-rotation approximation, the DRBBH metric \eqref{izr3n} can be approximated to the static bumblebee black hole~\cite{Casana:2017jkc}, which is given by 
\begin{equation}
ds^{2}\approx-F(r)dt^{2}+\frac{1}{g(r)}dr^{2}+r^2\left(d\theta^2 + \sin^2 \theta d\varphi^2\right), \label{iz2}
\end{equation}
where we recall that $F(r)=1-\frac{2M}{r}$ and 
\begin{equation}
\begin{aligned}
g(r)&=(\ell+1)\bigg(1-\frac{2M}{r}\bigg). \label{iz4}
\end{aligned}
\end{equation}

One can separate the scalar wave Equation \eqref{iz1} by considering the following ansatz: 
\begin{equation}
\Psi=R(r)Y_{l m}(\vartheta) e^{i m \phi} e^{-i \omega t} ,
\end{equation}
where $R(r)$ is nothing but the radial function and $Y_{l m}$ denotes the spherical harmonics. By employing the tortoise coordinate transformation $
r_{*}=\int \frac{dr}{\sqrt{f(r)g(r)}}$,  Equation \eqref{iz1} reduces to a one-dimensional Schr\"{o}dinger-like wave equation:
\begin{equation}
\frac{d^{2} R\left(r_{*}\right)}{d r_{*}^{2}}+\left(\omega^{2}-V\left(r_{*}\right)\right) R\left(r_{*}\right)=0,
\end{equation}
where $\omega=\omega_{R}-i \omega_{I}$ is a complex quasinormal mode frequency and the effective potential read is given by~\cite{Leung:1999iq}
\begin{align} \label{izpot}
V(r)&=H(r)\left(\frac{H^{\prime}(r)}{r}-\frac{l(l+1)}{r^{2}}\right),
\end{align}
where $H(r)=\sqrt{f(r)g(r)}$; recall that $l$ is the angular quantum number, and the prime symbol denotes the derivative with respect to $r$. Hence, one obtains 
\begin{align}\label{izpot1}
V(r)=&\frac{(r-2 M)\left(\frac{2 M}{r^3 \sqrt{1+\ell}}+\frac{l(l+1)}{r^2}\right)}{r \sqrt{1+\ell}}.
\end{align}

\begin{figure}[!ht]
    \includegraphics[scale=0.6]{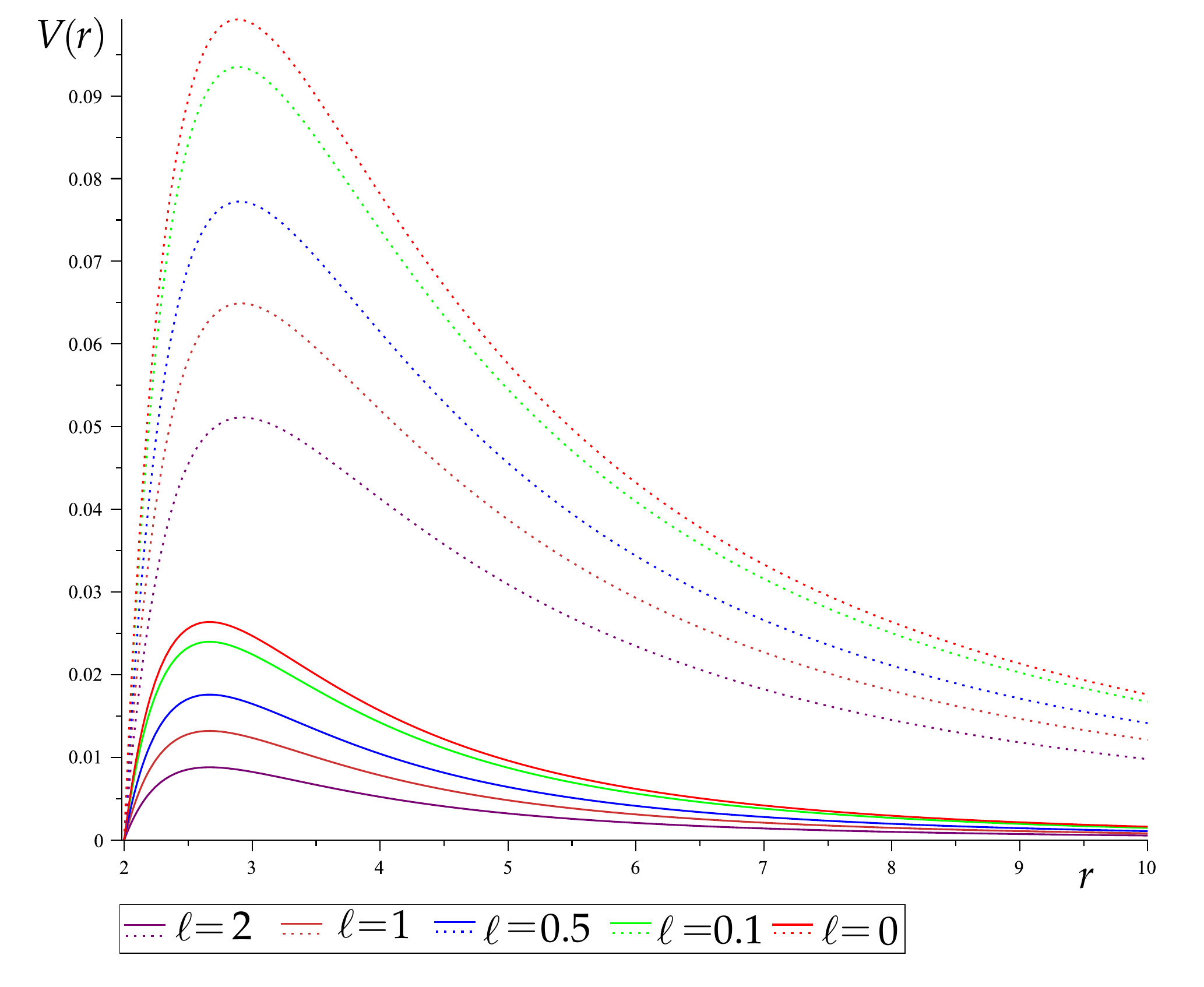}
    \caption{Plots of effective potential \eqref{izpot1} for massless spin-$0$ waves propagating in the extremely slow-spinning DRBBH \eqref{iz2}. The physical parameters are chosen as $M=1$ and $a\approx0$. The curves are labeled by the values of $l$: $l=0$ curves are represented by solid lines; dotted curves stand for $l=1$.}
    \label{test1}
\end{figure}

The behaviors of the effective potential for the massless bosonic waves propagating in the extremely slow-spinning DRBBH geometry \eqref{iz2} are depicted in Figure~\ref{test1}. It is clear from the figure that increasing the bumblebee parameter $\ell$ decreases the height of the potential barrier, while increasing the angular quantum number $l$ increases the apex of the barrier. 
In Reference~\cite{Cardoso:2008bp}, it was noted that the parameters governing the unstable circular null geodesics around any stationary spherically symmetric and asymptotically flat black holes---such as the angular velocity $\Omega_{C}$ and the principal Lyapunov exponent $\lambda_{\mathrm{Lexp}}$---have a remarkable correspondence with the QNMs emitted by the black hole in the eikonal (\mbox{i.e., short} wavelengths or high $l$ number) part of its spectrum, as discussed in Reference \cite{Khlebnikov:2007ii}. We now employ the WKB approximation method~\cite{Daghigh:2011ty} along with the unstable circular null geodesic method~\cite{Decanini:2010fz} to determine the frequency of the quasinormal mode. Specifically, in the large-$l$ limit, it was shown that the eikonal QNM frequencies are given by~\cite{Cardoso:2008bp}
\begin{equation}
\omega_{l \gg 1}=l \Omega_{C}-\frac{i}{2}\left(2n+1\right)\left|\lambda_{\mathrm{Lexp}}\right|, \end{equation}
where $n= 0, 1, 2, \dots$ is the overtone number, $\lambda_{\mathrm{Lexp}}$ denotes the Lyapunov exponent, and $\Omega_{C}$ represents the angular velocity at the unstable null geodesics, which is given by~\cite{Ovgun:2021ttv,Giri:2022zhf}
\begin{equation}
\Omega_{C}=\frac{\sqrt{H\left(r_{ps}\right)}}{r_{ps}}, 
\end{equation} 
where $r_{ps}$ is the radius of the photon sphere, which can be calculated by finding the largest root of this relation:
\begin{equation}
\frac{H\left(r_{ps}\right)}{H^{\prime}\left(r_{ps}\right)}=\frac{-2 M r_{ps}+r_{ps}^2}{2 M}=\frac{r_{ps}}{2}.
\end{equation}

After making a straightforward calculation, one can easily find that $r_{ps}=3M$. The expression of 
 $\lambda_{Lexp}$ can be derived as follows~\cite{Cardoso:2008bp}
\begin{equation}
\lambda_{\mathrm{Lexp}}=\sqrt{H\left(r_{ps}\right)\left[\frac{H\left(r_{ps}\right)}{r_{ps}^{2}}-\frac{1}{2}H^{\prime \prime}(r_{ps})\right]}.
\end{equation}

Table 1 \footnote{ The Table 1 is available in the published version}  illustrates the behaviors of the QNMs under varying values of the bumblebee parameter $\ell$, angular quantum number $l$, and overtone number $n$. Specifically, we observe that as $\ell$ increases, both the real and imaginary parts of the QNMs change. The real part ($Re(\omega)$) corresponds to the frequency of oscillations, while the imaginary part ($Im(\omega)$) dictates the rate at which the oscillations decay. Interestingly, all of the QNMs shown in Table 1 have negative imaginary parts, indicating that they are stable modes. In short, increasing the value of $\ell$ results in a decrease in the frequency $\omega$ of the scalar perturbations, which in turn causes the oscillations to decay more slowly. In other words, the higher values of $\ell$ correspond to longer-lived oscillations. Moreover, one can also observe that the increments in the $l$ and $n$-values increase the values of $Re(\omega)$ and $Im(\omega)$, respectively.

\section{Conclusions} \label{sec4}
In this paper, we presented a comprehensive discussion on gravitational lensing, GFs, and QNMs in RBBH spacetime. The study has provided impressive results in four dimensions when considering general relativity coupled with the bumblebee theory. After introducing some physical features of the SRBBH, which was proven to be the more physically correct metric version of RBBH \eqref{izr3}, to compute the gravitational lensing, we employed the RIM. The null geodesics and spherical 
photon orbit conditions were discussed to describe the gravitational lensing of SRBBH. By adjusting the Lorentz-violating parameter $\ell$ and modeling real stars to the SRBBH spacetime, gravitational lensing was depicted and its usual behavior was explored. We found that the presence of the Lorentz-violating parameter $\ell$ affects the bending angle of light moving in the SRBBH geometry, which increases with an increase in the Lorentz-violating parameter $\ell$. In summary, this result provides compelling evidence for indirectly detecting the existence of the Lorentz-violating parameter { $\ell$ and is, thus, a confirmation} of the bumblebee gravity theory.

In a further analysis, we discussed the scattering by using massive bosonic fields via the Klein--Gordon equation in SRBBH. In the sequel, we computed the GFs of the SRBBH by applying the semi-analytical bounds method. It has been found that the GFs of the SRBBH decrease with the increasing Lorentz-violating parameter $\ell$.  We  then extended our analysis to determine the QNM frequencies of the DRBBH. To this end, the null geodesics and spherical photon orbit conditions were discussed in order to employ the unstable circular null geodesic method using the Lyapunov exponents. After obtaining the one-dimensional Schr\"{o}dinger-like wave equation from the radial part of the Klein--Gordon equation, we illustrated the obtained effective potentials in Figure~\ref{test1} by varying the bumblebee parameter $\ell$, overtone number $n$, and angular quantum number $l$. We found that the existence of the Lorentz-violating parameter $\ell$ affects the potential barrier in such a manner that the peak values of the potentials reduce with the increasing bumblebee parameter $\ell$. Finally, we explored the QNMs
in the large $l$-limit, which consists of the imaginary (decaying) and real (oscillatory) parts. Our results show that both oscillatory and decaying sectors of the QNMs decrease with an increase in the magnitude of the 
Lorentz-violating parameter $\ell$. 

Our work can be extended to Kerr-like RBBH spacetimes for polarized light~\cite{Chen:2020qyp} as it opens up a new avenue 
to understand the bumblebee vector field with LSB and its interaction with the electromagnetic fields. Hence, the study of gravitational lensing, GFs, and QNMs in the effective bumblebee gravity metric for polarized light will be a natural extension of our work in the near future. Another possible future problem of bumblebee gravity in black hole physics is that it could lead to the formation of exotic objects known as ``gravastars''~\cite{Bhar:2021oag}. These objects are believed to be made up of a type of dark energy that can counteract the effects of gravity and prevent the formation of a singularity at the center of a black hole. While gravastars are still hypothetical, they could have important implications for our understanding of black hole physics and the nature of dark energy. We will also keep this issue among our possible future works.

\end{document}